# Cavity Induced Extraordinary Optical Transmission and Active Modulation with Graphene

Yifei Zhang[1#][*], Baoqing Zhang[1#], Mingming Feng[1], Haotian Ling[1], Xijian Zhang[1], Yiming Wang[1], Xiaomu Wang[2], Qingpu Wang[1], and Aimin Song[1,3][*]

[1]*Shandong Technology Center of Nanodevices and Integration, School of Microelectronics, Shandong University, Jinan, 250100, China*
[2]*National Laboratory of Solid State Microstructures, School of Electronic Science and Engineering, Nanjing University, Nanjing, 210023, China*
[3]*Department of Electrical and Electronic Engineering, University of Manchester, Manchester, M13 9PL, United Kingdom*

[*]Author to whom correspondence should be addressed. E-mail: yifeizhang@sdu.edu.cn.
[*]Author to whom correspondence should be addressed. E-mail: A.Song@manchester.ac.uk

[#]Authors contribute equally to this work

## Abstract

Extraordinary optical transmission (EOT) is a phenomenon of exceptional light transmission through a metallic film with hole arrays enhanced by surface plasmon (SP) resonance, which stimulates renewed research hotspots in metamaterials, subwavelength optics, and plasmonics. Below the frequency of the first order SP mode, $f_{pl0}$, the metallic film typically shows strong reflection and no EOT. Here, we report an unusual EOT phenomenon below $f_{pl0}$, i.e., beyond the long-held spectral boundary of classic EOTs. It is induced by a novel bound surface state in a Fabry–Pérot (F-P) cavity comprising a holey gold film and a silicon-air interface. By tailoring the cavity length, EOT phenomenon has been pushed deep into the sub-wavelength region by a factor of as large as 20%, and EOT frequency comb with cavity function has been achieved. Due to the enhanced slow-wave effect as the frequency approaches $f_{pl0}$, the cavity induced EOT gradually merges with the first order SP EOT. Distinguishing from the classic EOT phenomenon, no transmission zero is found between these two EOTs, which dramatically broadens the EOT bandwidth by a factor of 10 at terahertz (THz) frequencies. Furthermore, the

EOT transmittance is actively modulated with graphene, achieving a large modulation range from 0.5 to 0.25 under a sub-volt bias from -0.3 to 0.5 V at 500 GHz. To the best of the authors' knowledge, both the modulation range and the low bias are among the best for active EOT devices with graphene to date. Such a structure provides a new strategy for miniaturizing sensing devices, high-power sources, and broadband photonics as well as their active control in the THz regime.

## 1. Introduction

A metallic film perforated with subwavelength hole array gives rise to unexpectedly enhanced transmission at particular frequencies, which is known as the extraordinary optical transmission (EOT) [1-3]. The enhanced transmission efficiencies exceed unity, which hints that the subwavelength holes boost the wave transmission collectively with a transmittance larger than the hole area ratio [4-6]. Since Ebbesen et al. discovered this phenomenon in metallic films with circular holes [1], exhaustive theoretical and experimental explorations have been done to unveil its essence, i.e., how small the holes are, and how do the holes work [2-9]. Currently, a wide consensus for the physical mechanism of EOT is that two surface plasmon (SP) modes on the top and bottom metal surfaces weakly couple between themselves through the holes [5, 6]. Below the frequency of the first order SP mode, the metal surface shows little diffraction and, thus, light transmission [1, 5]. Surface plasmons are electromagnetic surface excitations trapped at a metallic surface through their interaction with the metal free electrons, which were originally reported at optical frequencies [10, 11]. In 2004, Pendry et al. established a close connection between optical SPs and terahertz (THz) surface modes on the perforated metal films by an effective permittivity of the same plasma form, revealing the SP nature of THz EOTs [10].

In longstanding theory, EOT typically has a sequence of transmission peaks isolated by transmission zeros in the spectrum, where standing SP modes dominate on the metal surfaces [4]. These SP modes offer strongly enhanced local field and, thus, high sensitivity for miniaturized sensing and surface-enhanced spectroscopy, such as fluorescence and Raman [4-6]. In this regard, the first order SP mode eventually sets the miniaturization limit in the spectrum for EOTs. Additionally, although the high transmission efficiency of EOTs blooms numerous stand-alone photonic devices [5, 6], e.g., filters, polarizers, and planar lenses, these devices typically show narrow bandwidth due to the plasmonic resonance as well as the presence of transmission zeros [12]. Their half-power bandwidth is merely several percent due to the negligible metal loss at THz and microwave frequencies [13-15], which significantly constrains their applications in the rising THz systems with large bandwidth consumption[16], such as ultrahigh-speed communication and passive imaging.

To provide programmable functions for the holey metal film with fixed optical properties, superconductors [14], $VO_2$ [17], and graphene [15, 18-20] have been investigated to dynamically modulate EOTs recently. Superconductor and $VO_2$ are thermally controlled materials so that their modulation speed is very limited. In contrast, graphene with high electron mobility allows for fast modulation up to several gigahertz [21]. However, the achieved amplitude modulation ranges are limited for EOTs, typically less than 20% [15, 18-20]. Moreover, the applied bias to graphene ranges from several volts to several hundred volts [23], which is not compatible with the standard silicon electronic technologies.

In this work, an unprecedented EOT phenomenon well below the frequency of the first order SP resonance ($f_{pl0}$) is discovered in a metal film with periodic square holes on a silicon substrate in the THz regime. The metal film and the silicon-air interface compose a weakly coupled Fabry–Pérot (F-P) cavity, which produces a novel hybrid bound surface state for the new EOTs and, thus, unlocks the spectral miniaturization limit for EOT. A frequency comb with enhanced transmission and cavity function is obtained below $f_{pl0}$. We note that the F-P cavity here is completely different from the F-P cavity in the waveguide-mode analysis of classic EOTs [5]. Furthermore, the cavity induced EOT gradually merges with the first order SP EOT as its frequency approaches $f_{pl0}$, showing no transmission zero between them. By carefully designing the cavity length, the two EOTs form an unusual broadband EOT phenomenon, whose half-power bandwidth is around ten times larger than the first order SP EOT. Finally, active modulation of the cavity induced EOT has been demonstrated by using graphene at a low bias from -0.3 to 0.5 V. A large amplitude modulation range is achieved from 0.5 to 0.25, which to our knowledge reaches the best for graphene-based active EOTs.

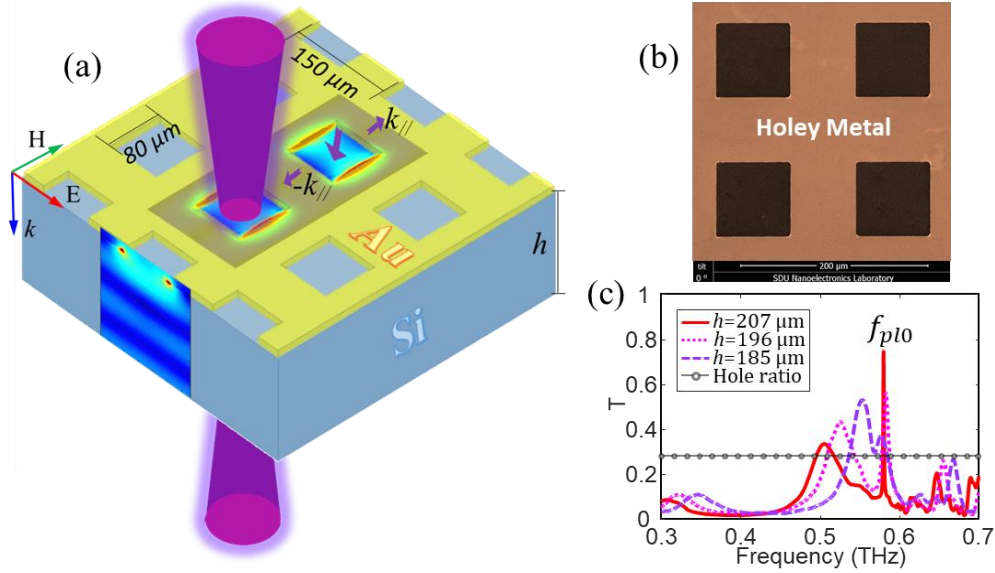

Fig. 1. Metallic film perforated with periodic square holes on a p-type silicon substrate. (a) Field distribution at the F-P resonant frequency and experimental configuration for wave transmission. (b) SEM image of a fabricated holey film. (c) Transmittances of the cavity induced EOTs with various substrate thickness. The new EOTs with a peak value larger than the hole area ratio (the gray circles) are well below the first order SP resonance ($f_{pl0}$ = 583 GHz) and enlarge as their frequencies approach $f_{pl0}$.

## 2. Experiment

Figure 1(a) illustrates a metal film perforated with square holes and its experimental configuration for wave transmission. The square holes with a size of 80 μm and a lattice period of 150 μm were fabricated on silicon substrates with various thickness, whose scanning electron microscope (SEM) image is shown in Fig. 1(b). The metal thickness is 0.3 μm, and the silicon thickness is denoted as $h$. The transmittances characterized using Toptica THz frequency-domain spectroscopy are shown in Fig. 1(c). The first order SP EOT lies at $f_{pl0}$ =583 GHz, and the high-order EOTs are found above 600 GHz, which are isolated by the transmission zeros. Remarkably, an unusually enhanced transmittance is observed well below $f_{pl0}$. The transmittances for $h$ =207, 196, and 185 μm are 0.336, 0.433, and 0.53 at 503, 525, and 550 GHz, respectively, all of which are larger than the area ratio of holes. The corresponding transmission efficiencies are 1.18, 1.52, and 1.89, respectively, revealing that the periodic holes enhance the light transmission collectively, i.e., the same function as the classic EOTs. It should be noted that the enhanced transmission overlaps with the first order SP EOT in the spectrum at $h = 185$ μm, inducing a broadband phenomenon with two transmission peaks and little transmission zeros between the peaks. The half-power bandwidth is 10%, i.e., approximately seven times larger than the first order SP EOT at $f_{pl0}$.

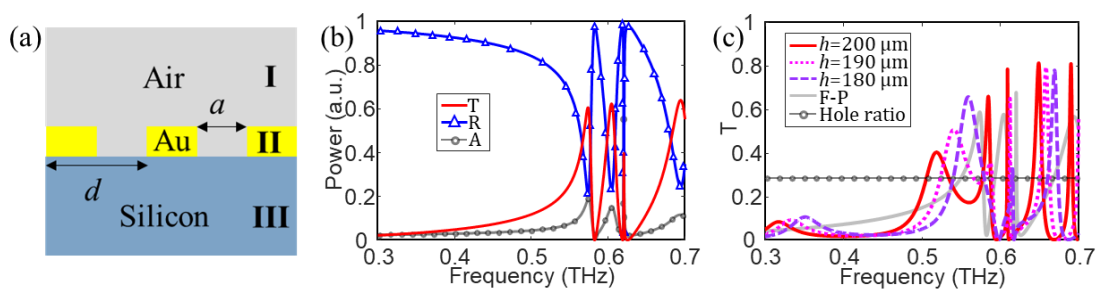

Fig. 2. (a) Canonical model of a holey metallic film sandwiched by semi-infinite air and silicon layers. (b) Simulated transmittance, reflectance, and absorption. The first order SP mode is at 575 GHz. (c) Simulated transmittance of the holey metal on a silicon substrate with finite thickness. The dotted line is the calculated F-P interference, which is smaller than the cavity induced EOTs.

## 3. Theory and Analysis

Intuitively, people may think this phenomenon is the cavity-induced transparency due to the F-P interference. To unveil the hidden mechanism, our story begins with a classic model of a holey metal film sandwiched by two semi-infinite media in Fig. 2(a) [5]. Light incidents normally from region I (air) and radiates to region III (silicon). Region II represents the perforated metal film with dispersive dielectric function $\varepsilon_m(\omega)$. The hole size and lattice period are denoted as *a* and *d*, respectively. Figure 2(b) illustrates the simulated transmittance, reflectance, and absorption. Two EOT peaks with Fano line-shape are at 575 and 605 GHz, respectively, and a transmission zero lies between these two peaks, showing good consistency with the reported works [4, 5]. The electric field distributions in Supplementary Information reveal clear SP properties for both EOT peaks, and the absorption also reaches peak value at the EOT frequencies due to SP attenuation. The wavelength of the first order SP mode is estimated as $\lambda = 2\sqrt{\varepsilon_r}a$ [5, 11], and the first transmission zero is found at $\lambda = \sqrt{\varepsilon_r}d$, where $\varepsilon_r$ is the effective permittivity of the substrate [2]. Note that the transmission efficiency is calculated as 2.34 and 2.38 at 575 and 605 GHz, respectively, and the half-power bandwidth of the first order SP mode at $f_{pl0}$ =575 GHz is approximately 1.4%. Below $f_{pl0}$, the subwavelength holes show little transmission and strong reflection.

Next, we consider the complex model with finite silicon layer in Fig. 1(a), where region III consists of a silicon substrate with a finite thickness and semi-infinite air. The simulated transmittances are shown in Fig. 2(c), where the classic SP EOTs slightly blue shifts due to the finite silicon thickness. A series of enhanced transmission peaks below $f_{pl0}$ are found in the spectrum, whose wavelengths are related to the F-P resonances by

$$2\sqrt{\varepsilon_r}h + \frac{\Delta\varphi}{2\pi}\lambda = n\lambda, \qquad (1)$$

where $\lambda$ is the wavelength in free space, $\Delta\varphi$ is the phase shift term due to the perforated metal film, and *n* is an integer. These peaks increase significantly as their frequencies approaches $f_{pl0}$. For *h* = 200 μm, the transmittances for the second order and third order F-P resonances are 0.08 and 0.4 at 317 and 519 GHz, respectively, whose transmission efficiencies are calculated as 0.28 and 1.48. The first order F-P resonance at ~110 GHz is too weak to be considered here. It should be noted that the third order cavity-induced transmittances are much stronger than the F-P interference. Taking 519 GHz for instance, the F-P interference is calculated as merely 0.18 according to Supplementary Information, and the enhanced transmittance is as high as 0.4 at 519 GHz, i.e., 2.2 times larger than the F-P interference. This phenomenon implies that the periodic holes together enhance the wave transmission, which is the key nature for EOTs. Here, we conclude this phenomenon is a new kind of EOTs below $f_{pl0}$. The EOT with a transmission efficiency larger than unity can be pushed down to around 470 GHz, as shown in Fig. 3(a), which reveals a red-shift factor of as large as 20%. Notably, a broadband EOT composing two transmission peaks can be tailored by varying the substrate thickness *h*. The maximum bandwidth is 13% at *h*=190 um, i.e., one order larger than that of the first order SP EOT, as discussed in Supplementary Information. Furthermore, a EOT frequency comb is designed below $f_{pl0}$, as shown in Fig. S-4(d) of Supplementary Information, whose F-P cavity function in the substrate may enable high-power laser applications in the THz regime.

Typically, a metal film with square hole array and a metal surface covering a layer of dielectric can bind a bound surface state [10]. The former plays a key role in the classic SP EOTs, whose dispersion relationship is

$$k_{//}^2 c_0^2 = (2\pi f)^2 + \frac{1}{f_{pl0}^2 - f^2}\frac{256 a^4 f^4}{\pi^2 d^4}, \quad (2)$$

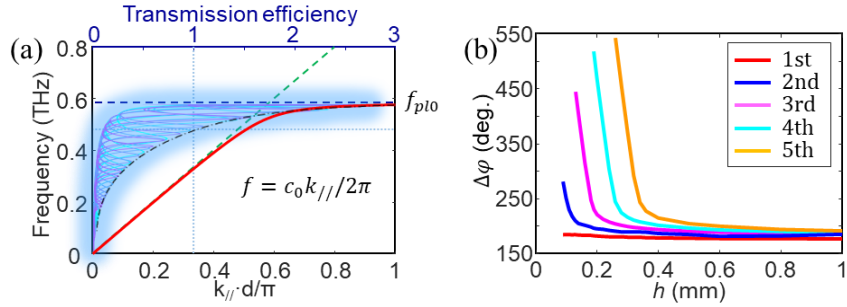

Fig. 3. (a) Dispersion relation of surface plasmons on the holey metal (the red line). The green dashed line depicts a linear curve for the light. The shadowed curves are the transmission efficiencies of the cavity-induced EOTs with various F-P resonant orders and substrate thickness. They increase from unity to 2.75 as the SP line deviates from the light line. (b) The phase item $\Delta\varphi$ induced by SP modes with various resonant orders. As the thickness $h$ decreases, the F-P resonance approaches $f_{pl0}$, and $\Delta\varphi$ exponentially increases due to the enhanced slow-wave effect.

where $k_{//}$ is the wavenumber of SPs [10]. As depicted in Fig. 3(a), the calculated dispersion relation (the red solid line) gradually deviates from the light line (the green dashed line) and approaches the effective plasma frequency $f_{pl0}$ (the first order SP mode), revealing enhanced slow-wave effect [5]. It is interesting to note that the frequency of the new EOTs gradually stops at $f_{pl0}$ and merges with the first order SP EOT as the substrate gets thinner (and the frequency increases), the trend of which coincides with the dispersion relation, see the shadowed lines in Fig. 3(a). Simultaneously, the EOT transmittances enhance significantly as the frequency increases, which also coincides with the dispersion relation. This phenomenon can be described by an unusual hybrid bound state comprising SPs on the holey metal and F-

P resonance. The holey metal induces extra phase term $\Delta\varphi$ for the F-P cavity, which is depicted as

$$\frac{\Delta\varphi}{\Delta l} = \frac{2\pi}{\lambda} = k_{//}, \tag{3}$$

where $\Delta l$ is the length for $\Delta\varphi$ on the holey metal film. According to Equation (1), $\Delta\varphi$ can be derived from the F-P resonant frequencies, which is plotted in Fig. 3(b). As the thickness $h$ gets smaller, and the F-P resonance approaches $f_{pl0}$ (see Fig. S-3 in Supplementary Information), $\Delta\varphi$ ramps fast due to the enhanced slow-wave effect, and so does $k_{//}$. As the frequency departs from $f_{pl0}$, the surface state releases bound strength as the canonical optical SPs, and the cavity enhanced transmission functions more like an F-P interference with a phase shift of 180° at the metal reflector, see Fig. S-3(c) in Supplementary Information.

## 4. Active Modulation

As a promising candidate for active modulation of metamaterials and plasmonics [18-24], graphene is utilized to actively modulate the cavity induced EOT in this work. After the fabrication of the holey metal film, a chemical vapor deposition (CVD) graphene was transferred onto the film, and then solution processed poly (styrenesulfonic acid sodium salt) (PSSNa) was spin-coated on the graphene with a thickness of 70 μm for top-gating. The PSSNa with a dielectric constant larger than one broadens the EOT bandwidth up to 22%, and enhances the cavity-induced EOT and the first order SP EOT to 0.896 and 0.863, respectively, which reduce to 0.5 and 0.17 with the presence of graphene at a bias of -0.3 V, as illustrated in Fig. 4(a). The hybrid bound state shows much smaller graphene attenuation than the classic one so that the transmittance of the cavity-induced EOT is much stronger. Furthermore, adding a thin dielectric layer on top of the holey metal film can further push EOTs to 366 GHz, as shown in Fig. S-4(c) of Supplementary Information, which approximately corresponds to a red-shift factor of 38% with respect to the first order SP EOT.

The spectral transmittances at various bias are illustrated in Fig. 4(a), and the absolute variation values referenced to the transmittance at 0.5 V are illustrated in Fig. 4(b). The transmittance for the cavity induced EOT enlarges from 0.42 to 0.5 under a bias sweeping from -0.7 to -0.3 V and reduces from 0.5 to 0.25 under a bias sweeping from -0.3 to 0.5 V, showing reasonable agreement with the simulation in Fig. S-5 of Supplementary Information. Meanwhile, the transmittance of the first order SP EOT is also modulated from 0.17 to 0.07 at 578 GHz. Figure 4(c) illustrates the corresponding modulation depths for both EOTs, which are larger than 100% at -0.3 V.

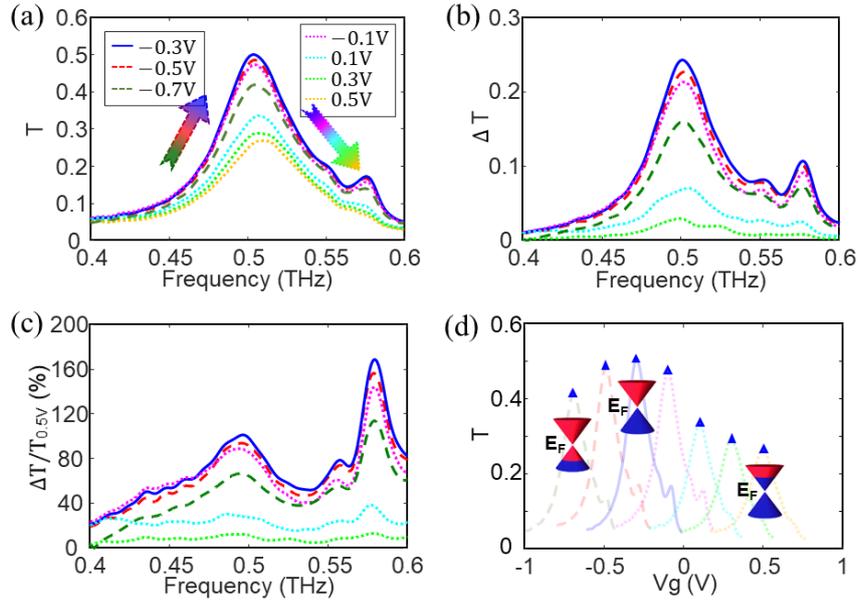

Fig. 4. (a) Switching spectral performance of the active EOT with PSSNa ($h$ =207 μm). (b) Transmittance variation using the transmittance at 0.5 V ($T_{0.5V}$) as a reference. The peak value is 0.25. (c) Modulation depth with respect to $T_{0.5V}$. (d) Switching performance as a function of gate bias. The triangles depict the transmittance peaks.

Figure 4(d) depicts the transmittance as a function of the applied gate bias. The transmittance reaches peak value at -0.3 V, which is consistent with the Dirac point obtained from the current-voltage measurement of a graphene transistor top-gated by PSSNa on the same substrate. The charge carriers switch from electrons to holes with a quasi-symmetric shape as the gate voltage sweeps from larger than the Dirac point to smaller than the Dirac point. The electrons show better modulation capability for THz waves than the holes. Here, it is worthy to mention that the modulation range of our device is among the state-of-art for active EOTs with graphene, and the applied bias is among the lowest ones for graphene tuned metamaterials [18-24].

## 5. Perspective

Recently, THz technologies are rising for an increasingly wide variety of applications [16], such as biosensing and medical diagnosis [25], ultrahigh-speed communication [26], and quantum optics [27], etc. For the past two decades, EOT has bloomed numerous studies in metamaterials, plasmonics, and quantum optics for two major applications: sensing and photonic devices [4-6]. In this regard, the THz cavity induced EOT provides a novel approach with more subwavelength holes for sensing molecules and chemicals as well as for studying THz quantum devices [16, 27]. The achieved EOT frequency comb with cavity function may be attractive for THz quantum cascaded lasers [6, 28]. Furthermore, the unique broadband property of the cavity induced EOTs enables stand-alone filters, polarizers, and lenses with considerable bandwidth for the advanced THz communication and imaging applications. Finally, the reported approach of active modulation makes the programmable EOT devices compatible with typical silicon complementary metal–oxide–semiconductor (CMOS) bias voltages.

## 6. Methods

**Sample fabrication.** The holey metal films were fabricated with standard ultraviolet photolithography (AZ6130 photoresist), E-beam evaporation (HHV Auto500), and lift-off process. To achieve various substrate thickness, the silicon substrates were etched by using inductively coupled plasma (ICP) etching system (Oxford PlasmaPro 100 Cobra). The etch rate is 2.8 μm/min under a RF power of 30W, an ICP power of 825 W, and a mixed gas of $O_2$, $SF_6$ and Ar (2 sccm, 26 sccm and 20 sccm).

**Ion gel preparation.** PSSNa, D-sorbitol, glycerol and DI water (with a weight ratio of 40, 10, 10, and 40%) were mixed with magnetic stirring. After stirring for two hours at room temperature, the ion gel was spin-coated onto the metallic film with a spinning rate of 1000 rpm for 1 min.

**Characterization.** The transmittances were characterized using Toptica TeraScan 1550 THz frequency-domain spectroscopy (FDS) at room temperature, which are normalized to the water vapor absorption lines with stable humidity. The incident terahertz wave was generated by using two continuous wave lasers with differential frequency method, and focused with a beam radius of around 2 mm in a four-mirror reflection system. The spectral resolution is as small as 10 MHz.

**Simulation.** The 3-D models of the EOT structures are simulated with Master and Slave boundaries, i.e., a kind of periodic boundary, in Ansys High Frequency Structural Simulator (HFSS).

## References:

1. Ebbesen, T. et al. Extraordinary optical transmission through sub-wavelength hole arrays. *Nature* 391, 667-669 (1998).
2. Liu, H. & Lalanne, P. Microscopic theory of the extraordinary optical transmission. *Nature* 452, 728-731 (2008).
3. van Beijnum, F. et al. Quasi-cylindrical wave contribution in experiments on extraordinary optical transmission. *Nature* 492, 411-414 (2012).
4. Genet, C. & Ebbesen, T. W. Light in tiny holes. *Nature* 445, 39-46 (2007).
5. Garcia-Vidal, F. J. et al. Light passing through subwavelength apertures. *Rev. Mod. Phys.* 82, 729-787 (2010).
6. Rodrigo, S., Fernando, L. P. & Luis, M. M. Extraordinary optical transmission: fundamentals and applications. *Proc. IEEE* 104, 2288-2306 (2016).
7. Zhong, X. et al. Waveguide and plasmonic absorption-induced transparency. *ACS Nano* 10, 4570-4578 (2016).
8. Chen, Z. et al. Enhanced extraordinary optical transmission and refractive-index sensing sensitivity in tapered plasmonic nanohole arrays. *Nanotechnology* 30, 335201 (2019).
9. Lee, J. et al. Extraordinary optical transmission and second harmonic generation in sub–10-nm plasmonic coaxial aperture. *Nanophotonics* 9, 3295-3302 (2020).
10. Pendry, J. B. et al. Mimicking surface plasmons with structured surfaces. *Science* 305, 847-848 (2004).
11. Hibbins, A.P. et al. Experimental verification of designer surface plasmons. *Science* 308, 670–672 (2005).
12. Alu`, A. et al. Plasmonic Brewster angle: broadband extraordinary transmission through optical gratings. *Phys. Rev. Lett.* 106, 123902 (2011).
13. Gong, S. et al. Ultra-extraordinary optical transmission induced by cascade coupling of surface plasmon polaritons in composite graphene-dielectric stack. *Opt. Express* 28 (12), 30502 (2020).
14. Wu, J. B. et al. The dynamic process and microscopic mechanism of extraordinary terahertz transmission through perforated superconducting films. *Sci. Rep.* 5, 1-12 (2015).
15. Shaltout, A. M. et al. Spatiotemporal light control with active metasurfaces. *Science* 364, eaat3100 (2019).
16. Sirtori, C. Terahertz race heats up. *Nat. Photonics* 15, 1-2 (2021).
17. Kang, T. et al. Large-scale, power-efficient Au/VO2 active metasurfaces for ultrafast optical modulation. *Nanophotonics* 10, 909-918 (2020).
18. Gao, Z. et al. Tunable extraordinary optical transmission with graphene in terahertz. *ACS*

*Omega* 6(44), 29746–29751 (2021).
19. Kim, S. et al. Electronically tunable extraordinary optical transmission in graphene plasmonic ribbons coupled to subwavelength metallic slit arrays. *Nat. Commun.* 7, 12323 (2016).
20. Barzegar-Parizi, S. & Khavasi, A. Tunable extraordinary transmission through a graphene-covered hole array: an analytical equivalent-circuit modeling approach. *Plasmonics* 14, 569-577 (2018).
21. Low, T. & Avouris, P. Graphene plasmonics for terahertz to mid-infrared applications. *ACS Nano* 8, 1086-1101 (2014).
22. Lee, S. H. et al. Switching terahertz waves with gate-controlled active graphene metamaterials. *Nature Mat.* 11, 936-941 (2012).
23. Ju, L. et al. Graphene plasmonics for tunable terahertz metamaterials. *Nature Nanotech.* 6, 630-634 (2011).
24. Wang, Z. et al. Recent progress in terahertz modulation using photonic structures based on two-dimensional materials. *InfoMat* 3, 1110-1133 (2021).
25. Wu, X. et al. Programmable terahertz chip-scale sensing interface with direct digital reconfiguration at sub-wavelength scales. *Nat. Commun.* 10, 2722 (2019).
26. Nagatsuma, T. et al. Advances in terahertz communications accelerated by photonics. *Nat. Photonics* 10, 371-379 (2016).
27. Kutas, M. et al. Terahertz quantum sensing. *Sci. Adv.* 6, eaaz8065 (2020).
28. Yu, N. et al. Designer spoof surface plasmon structures collimate terahertz laser beams. *Nature Mater*. 9, 730-735 (2010).